\documentclass[12pt,preprint]{aastex}
\usepackage{psfig,natbib}
\shortauthors{Bower et al.}
\shorttitle{Intrinsic Size of Sagittarius A*}
\begin{document}

\newcommand\degd{\ifmmode^{\circ}\!\!\!.\,\else$^{\circ}\!\!\!.\,$\fi}
\newcommand{\etal}{{\it et al.\ }}
\newcommand{\uv}{(u,v)}
\newcommand{\rdm}{{\rm\ rad\ m^{-2}}}
\newcommand{\msuny}{{\rm\ M_{\sun}\ y^{-1}}}
\newcommand{\mylesssim}{\stackrel{\scriptstyle <}{\scriptstyle \sim}}
\newcommand{\sci}{Science}


\title{The Intrinsic Size of Sagittarius A* from 0.35 cm to 6 cm}

\author{Geoffrey C. Bower\altaffilmark{1}, W.M. Goss\altaffilmark{2}, Heino Falcke\altaffilmark{3}, Donald C. Backer\altaffilmark{1}, Yoram Lithwick\altaffilmark{4}} 
\altaffiltext{1}{Astronomy Department \& Radio Astronomy Laboratory,
University of California, Berkeley, CA 94720; gbower,dbacker@astro.berkeley.edu}
\altaffiltext{2}{National Radio Astronomy Observatory, P.O. Box 0, Socorro NM
87801, U.S.A. ; mgoss@nrao.edu}
\altaffiltext{3}{ASTRON, Postbus 2, 7990 AA Dwingeloo, The Netherlands and Department 
of Astrophysics, Radboud Universiteit Nijmegen, Postbus 9010, 6500 GL Nijmegen, 
The Netherlands ; falcke@astron.nl}
\altaffiltext{4}{CITA, University of Toronto, 60 St. George Street,
Toronto, Ontario, M5S 3H8; yoram@cita.utoronto.ca}

\begin{abstract}

We present new high-resolution observations of Sagittarius A* at 
wavelengths of 17.4 to 23.8 cm with the Very Large Array in A configuration
with the Pie Town Very Long Baseline Array antenna.  We use the
measured sizes to calibrate the interstellar scattering law and find
that the major axis size of the scattering law is smaller by $\sim 6\%$
than previous estimates.  Using
the new scattering law, we are able to determine the intrinsic size of
Sgr A* at wavelengths from 0.35 cm to 6 cm using existing results from
the VLBA.  The new law increases the intrinsic size at 0.7 cm by $\sim 20\%$ and $< 5\%$ at
0.35 cm.  The intrinsic size is $13^{+7}_{-3}$ Schwarzschild radii at 
0.35 cm and is proportional to $\lambda^{\gamma}$, where $\gamma$ is in the
range 1.3 to 1.7.

\end{abstract}

\section{Introduction}

Imaging the radio emitting region surrounding the massive black hole in the
Galactic Center, Sagittarius A*, has been a goal since its discovery
\citep{1974ApJ...194..265B}.
Turbulent electrons along the line of sight to
Sgr A*, however, scatter radio wavelength photons and produce an image
that is an elliptical Gaussian with a major axis size of $\sim 0.5$ arcsec
at 20 cm and a $\lambda^2$ dependence
\citep{1978ApJ...222L...9B}.  Separating the effects of the
small intrinsic source from the effects of scattering has required 
observations at short wavelengths,
careful calibration, and the use of closure amplitude
techniques, which reduce
sensitivity but remove uncertainty due to calibration error
\citep[e.g.,][]{2004Sci...304..704,2005Natur.438...62S}.  These
efforts have recently resulted in the first robust detections of 
intrinsic structure in Sgr A* at wavelengths of 1.3 cm, 0.7 cm and 0.35 cm.
The intrinsic source has a size
that scales with $\lambda^{1.1}$ or $\lambda^{1.6}$ 
to a minimum of $\sim 10$ Schwarzschild
radii at 0.35 cm (assuming $M_{bh}=4 \times 10^6 M_\sun$ and $d=8$ kpc ;
\citet{2005ApJ...620..744G,2003ApJ...597L.121E}).

These detections of the intrinsic size of Sgr A* have a number of consequences.
The brightness temperature of $10^{10}$ K strongly excludes advection
dominated accretion flows \citep[ADAFs;][]{1998ApJ...492..554N}
 and Bondi-Hoyle accretion models \citep{1994ApJ...426..577M}.  These size
measurements, however,
 cannot differentiate between jet models \citep{1993A&A...278L...1F}, generic radiatively
inefficient accretion flows \citep{2000ApJ...539..809Q}, and hybrids of these models
\citep{2002A&A...383..854Y}.  This limitation is 
principally due to the limited sensitivity in the minor axis size of 
the scattering ellipse.  Coupled with measurements of the proper motion
of Sgr A* \citep{2004ApJ...616..872R}, the assumption that the 
black hole is smaller than
the emission region implies a lower limit to the mass density of the black
hole $\sim 10^5 M_\sun {\rm AU^{-3}}$, which  strongly excludes
alternative models for dark mass objects.

The scattering medium itself is a system of intense interest
\citep{1998ApJ...505..715L,2001ApJ...558..127B,2006ApJ...640L.159G}.
The $\lambda^2$ 
size dependence of Sgr A* is a strict consequence
of the strong scattering and the short projected baselines at the
distances of the scattering medium
\citep{1989MNRAS.238..963N}.

We present here detailed measurements of the scattering properties at 
wavelengths that range from 17.4 to 23.8 cm using the Very Large Array 
and the Pie Town Very Long Baseline Array antenna (\S 2).  The
addition of the PT antenna to the VLA A configuration improves the
East-West resolution by a factor of two.  
The resulting scattering law is smaller by $\sim 6\%$ than previous estimates.
The independent estimate of the scattering law enables us to measure
the intrinsic size of Sgr A* at a wavelength
as long as 6 cm (\S 3).  We discuss these results in (\S 4).

We wish to clarify here the numerous axes and steps involved in translating 
observations of Sgr A* into a measurement of the intrinsic size.  VLA 
observations of Sgr A* are obtained with a synthesized beam that is extended
North-South.  Deconvolution of the observed image with the synthesized beam
gives the apparent, scatter-broadened image of Sgr A*.  This image is 
predominantly a two-dimensional Gaussian with the major axis oriented 
$\sim 80$ degrees East of North.  Throughout
this paper when we refer to the major and minor axes, we refer to 
the orientation of the scattering Gaussian.  Finally, to obtain the intrinsic
image, we deconvolve the apparent image with a model of the scatter-broadened
image, which is determined from long wavelength apparent sizes.

\section{Observations and Analysis}

We obtained new observations of Sgr A* with the Very Large Array plus the
Pie Town VLBA antenna.  The VLA was in the A configuration.  Observations
were made on 1 and 4 October 2004 in eight separate bands
centered at 
25.2, 23.8, 23.2, 21.9, 20.9, 19.8, 18.0, and 17.5 cm.
Each band had 12.5 MHz bandwidth with 15 channels.  Results at 
25.2 and 21.9 cm
were corrupted by interference and we do not consider
these data any further.

We calibrated the absolute flux density with observations of 3C 286.
Corrections for atmospheric and instrumental amplitude and phase fluctuations
were made through self-calibration of frequent observations of the
compact source J1744-312.  We imaged Sgr A* using baselines longer than
$50 k\lambda$ and uniform weighting to suppress large-scale structure
(Figure~\ref{fig:vlapt}).

The presence of a radio
transient with a flux density of $\sim 30 $ mJy and a resolved morphology
at a separation of 2.7\arcsec\ South of Sgr A* precluded modeling
in the visibility and closure amplitude domains 
\citep{2005ApJ...633..218B}.  We previously showed that fitting 
in the image domain
provides results that are equivalent to fitting in the closure amplitude domain
at centimeter wavelengths, in the case where the difficulty of calibration 
and poor telescope performance are less critical
\citep{2001ApJ...558..127B,2004Sci...304..704}.  Long wavelength data 
obtained from the VLA meet these criteria better than any other data.
Accordingly, we fit sizes to Sgr A* in the image plane with a region
that excluded most of the transient flux density.  The effect of the transient
is primarily on the accuracy of 
the size in the minor axis of the scattering angle.
The synthesized beam ranges from $1.69 \times 0.56$ arcsec at 17.5 cm
to $2.36 \times 0.98$ arcsec at 23.8 cm,
oriented roughly North-South.  
Measured sizes ranged from $1.71 \times 0.67$ arcsec at 
17.5 cm and 
$2.43\times 1.22$ arcsec at 
23.8 cm.  Sgr A* is clearly resolved in both
axes but with considerably more accuracy in the East-West axis than an in
the North-South axis.  Fitting a point source to the data produced very
poor quality fits, while fitting an elliptical Gaussian produced a residual image
with no obvious systematic errors and an rms $\sim 2.5$ mJy/beam (Figure~\ref{fig:vlapt}).
This rms is a few times the rms $\sim 0.9$ mJy/beam 
determined far away from Sgr A*, possibly due to
the presence of confusing emission around Sgr A*.

We deconvolved the measured Gaussian sizes with the synthesized beam sizes
 to determine the true source parameters:  total flux density,
major axis, minor axis, and position angle.  We determined errors in
the parameters by calculating $\chi^2$ for a grid in the parameters 
surrounding the best-fit value.  These errors are the formal uncertainy
in the parameters and do not reflect the systematic errors, which
we discuss below.  Results are tabulated in Table~1.

In the case of the minor axis,
there is a clear trend of decreasing
size with decreasing wavelength, which indicates the
presence of systematic errors.  Marginal resolution of Sgr A* in the
North-South axis and confusion from the presence of the radio transient
due South of Sgr A* are the likely causes of this effect.
The major axis size, however, is not affected by the transient and
only weakly distorted by changes in the position angle; 
the best-fit solution for major axis size
changes by only 1\% with a $10^\circ$ change in position angle.  
We conclude that the results determined from the VLA+PT result
are accurate in the major axis but not in the minor axis.

We experimented with a range of imaging 
parameters to explore systematic effects on the deconvolved size of Sgr A*.
Weighting with a robustness parameter of 0, using super-uniform weighting,
and changing the minimum $\uv$ distance from 40 to 100 $k\lambda$ changed
the deconvolved major axis size by no more than 3\%.  Since the results are
strongly dependent on the PT antenna, we dropped random groups of 
5 baselines associated with PT, producing  1\% changes in the deconvolved size.
These errors are comparable to those found for other sources through VLA
observations \citep[e.g.,][]{1998ApJ...493..666T}.
Thus, our results are influenced by systematic imaging effects at a level 
of a factor of no more than a few.  As we discuss below, a factor of a few is 
consistent with the scatter in the measurements.

\section{The Scattering Size and The Intrinsic Size of Sgr A*}

In Figure~\ref{fig:results}, we plot the measured size of Sgr A* from the VLA+PT observations
and from Very 
Long Baseline Array results from 
\citet{2004Sci...304..704}
at wavelengths
of 6.01 cm to 0.67 cm.  We also include the VLBA result from 
\citet{2005Natur.438...62S}
at 0.35 cm.  The sizes are plotted normalized by wavelength squared.

We fit a power law of the form $\lambda^2$ to the major axis
sizes using the new VLA+PT results at 17.4 to 23.8 cm wavelength.  The
best-fit
value to the normalization of the scattering law is $1.309 \pm 0.017$ 
mas/cm$^2$.  The errors in these values are
determined from the scatter in the measurements.  The best-fit scattering
values are
plotted as a straight line in Figure~\ref{fig:results}.  
The major axis normalization is $\sim 6\%$
smaller than previous estimates.  None of our measurements are consistent with 
the previously determined major axis normalization of 1.39 mas/cm$^2$.
The best estimates of the minor axis scattering size and position angle
remain the results determined previously from VLBA observations
at wavelengths between 2 cm and 6 cm \citep[$0.64^{+0.04}_{-0.05}$ mas/cm$^2$
and $78^{+0.8}_{-1.0}$ deg;][]{2004Sci...304..704}.

The scatter in the VLA+PT
major axis sizes is much larger than the expectation of the statistical errors
for individual points.  The reduced $\chi^2_\nu \approx 7$ for the major 
axis, indicating that there are additional sources of error in the measurement of the
size that we are not including.  Dropping either the two highest frequency or two 
lowest frequency VLA+PT sizes did not significantly affect the reduced $\chi^2$ or the 
best-fit scattering law.   We also explored the effect of the inclusion of the shorter 
wavelength VLBA results on the major axis scattering size.  Inclusion of the 6 cm size 
decreases the reduced $\chi^2_\nu$ slightly but does not affect the scattering size significantly.
Inclusion of VLBA results at wavelengths of 3.6 cm and shorter, however, leads to a significant
increase in $\chi^2_\nu$ to 18.  That is, the major axis sizes at $\lambda <= 3.6$ cm 
are not consistent with the longer wavelength sizes and a $\lambda^2$ scattering law.
This inconsistency holds if we scale the VLA+PT errors
by a factor as large as 5, which reveals $\chi^2_\nu=1.8$.

The $\lambda^2$ dependence of the scattering law is strongly favored for
theoretical reasons.  The maximum baseline length projected to the scattering 
region is $b_{proj}=D_{scattering}/D_{SgrA} \times b_{max} \sim 1 km$, 
where $D_{SgrA}=8$ kpc is the distance to Sgr A*, $D_{scattering}\approx 100$ 
pc is the distance of the scattering region from Sgr A*, and 
$b_{max}\approx 70$ km is the maximum baseline between the VLA
and PT.  $b_{proj}$ is substantially smaller than the expected and measured 
inner scales ($b_{inner}$) for the power spectrum of turbulent 
fluctuations ($10^2$ to $10^{5.5}$ km; \citet{1994MNRAS.269...67W}).
This result holds for the long VLBA baselines involved in imaging at 
shorter wavelengths, as well, where $b_{proj} \sim 25$ km at 0.7 cm 
wavelength.  For the case of $b_{proj} << b_{inner}$, the resulting image 
is very heavily averaged and must be Gaussian in shape with 
size $\propto \lambda^2$
\citep{1989MNRAS.238..963N}.

This expectation of strong scattering is supported
by previous measurements of the shape of the image of Sgr A*.
Bower et al. (2004) showed 
that fitting the closure amplitudes of Sgr A* at 0.7 cm 
with a functional form for
the visibilities of $\propto e^{-b^{(\beta-2)}}$, where
$b$ is the baseline length,  revealed
$\beta=4.00 \pm 0.03$. That is, the best evidence
indicates that the image of Sgr A* is a Gaussian. 
Following scattering theory, 
where the size is proportional to $\lambda^\alpha$,  
then $\alpha=\beta-2=2.00 \pm 0.03$
\citep{1989MNRAS.238..963N}.  For the case of 
the VLA+PT data, we find $\alpha=1.98 \pm 0.11$, which is consistent
with the expectation of $\alpha=2$.

Nevertheless, if we
fit a single power-law to all of the VLA+PT and VLBA
data, we find $\alpha=1.96 \pm 0.01$ with a $\chi^2_\nu=5.5$.  
So, without the constraint of a $\lambda^2$ size for the scattering
law at some wavelength, the evidence for resolving 
an intrinsic size becomes marginal at any wavelength.
A fit of the size proportional to 
$\sqrt{ a^2\lambda^4 + b^2\lambda^{2\gamma}}$ 
produces an apparent size very similar to that of the single index power-law
fit but is not sufficiently constrained to set reasonable limits
on the parameters.  If we fix $\gamma$ and search for $a$ and $b$, we find
that $\chi^2$ for $\gamma=1$ is four times the value for $\gamma$ unconstrained, indicating
that $\gamma=1$ is strongly excluded.  Without the 
assumption that the second term is negligible for wavelengths longer than
6 cm, therefore, we cannot determine the scattering law or the intrinsic
size of Sgr A*.

A final caveat is required.  The scattering medium is dynamic.  The minimum
time scale for a change in the medium is the refractive time scale, which is 
$0.5 y (v/100 km/s) (\lambda/1 cm)^2$ for Sgr A*, where $v$ is the velocity of
the scattering material relative to Sgr A* and the Sun
\citep{1989MNRAS.238..963N}.  The data presented here were obtained in a 
span of roughly a decade.  The long-wavelength scattering properties
are very unlikely to change on this time scale.  However, 
at wavelengths as long as 4 cm, the refractive time scale is $<10$ year.
The many observations at 0.7 cm in this period, however, appear to produce a
source of stable size, despite a refractive time scale less than one year.
We conclude it is unlikely that the scattering size has changed significantly
over this period.

With the assumption that the scattering law is determined accurately
at wavelengths longer than 17 cm, we can determine the intrinsic size.  
We subtract in quadrature 
the scattering law size from the measured size (Table~2,
Figure~\ref{fig:intrinsic}).  We compute
the results for the best-fit major axis scattering law 
($b_{sc}=1.31 {\rm\ mas/cm^2}$), 
and $\pm 3\sigma$ of
the best-fit value.  For the best-fit case,
the intrinsic size is accurately determined from 0.35 cm to 3.6 cm.
Over this range, the intrinsic size is well-fit by a power-law
$\lambda^\gamma$, where $\gamma=1.6 \pm 0.1$.  For the smaller scattering 
size, we find a steeper power-law and measure the intrinsic size from 0.35 cm 
to 6 cm.  For the larger scattering size, we cannot measure the intrinsic size 
at wavelengths longer than
1.3 cm and find a shallower power-law index of $\gamma=1.3 \pm 0.2$.

If the intrinsic size power-law extends to $\lambda\sim 20$ cm, the 
contribution of the intrinsic size results in an increase of
the measured angular sizes by $\sim 1.5\%$.  This is comparable to the
error in the major axis scattering law and, therefore, negligible.

\section{Discussion}

We have measured the intrinsic size of Sgr A* from 0.35 to 3.6 cm.
At short wavelengths, the result is consistent with the conclusions
of recent efforts by 
\citet{2004Sci...304..704} and \citet{2005Natur.438...62S}.
The
size of Sgr A* at 0.35 cm is 13.3$^{+6.7}_{-3.1} R_s$, where $R_s=1.2 \times 10^{12}$
cm
is the Schwarzschild radius  for $M_{bh} = 4 \times 10^6 M_\sun$ and
the Galactic Center distance $d=8$ kpc.  This compact size confirms
tight restrictions on accretion models and black hole alternatives
previously claimed and stated in \S 1.

The wavelength dependence of the source size agrees with that found
by 
\citet{2004Sci...304..704} 
and is steeper than that found by \citet{2005Natur.438...62S},
who found $\gamma\approx 1.1$.  The steeper dependence indicates that
the brightness temperature decreases as $\lambda^{-1}$, assuming that
the size in the second dimension is proportional to the major axis size.
The peak brightness temperature at 0.35 cm is $\sim 10^{10}$ K for a
flux density of 1 Jy.
The power-law dependence of the size as a function of wavelength indicates
a stratified, smoothly varying emission region.

Detailed jet models for Sgr A* predict $\gamma \approx 1$ 
\citep{2000A&A...362..113F}.
Generalized jet models, however, allow a range of $\gamma$, depending
on the details of the magnetic field and particle energy density 
distributions 
\citep[e.g.,][]{1981ApJ...243..700K}.
A jet with $B \propto r^{-1}$,
electron density decreasing as $r^{-1}$, and optically thin power-law
index of 1 will show a size $\propto \lambda^{1.4}$.

\citet{2006ApJ...642L..45Y}
model the size of Sgr A* for a radiatively inefficient
accretion flow.  They fit sizes at 0.35 cm and 0.7 cm that are fit
with a power law of index $\gamma=1.1$.    Variations in the
nonthermal electron distribution or deviations from equipartition, however, 
could alter $\gamma$ in their model.

The critical remaining observational goals for understanding the image
of the radio emission of Sgr A* are a measurement of the two-dimensional
size and detection of structural variability.  The simple one-dimensional
deconvolution that we have performed here only gives schematic information
on the size of Sgr A*.  With a more accurate 
two-dimensional scattering model, future analysis will directly
compare the observed image with
non-Gaussian source models convolved with the imaging 
and scattering constraints.  Astrometric observations 
may indicate a shift in the centroid of the image with frequency.  
A heterogeneous jet will exhibit such a shift due to changing location
of the optically thick surface of the source 
\citep{1981ApJ...243..700K}.

At mm and submm wavelengths, the gravity of the 
black hole will distort the image \citep{2000ApJ...528L..13F,2005MNRAS.363..353B}.
Detailed knowledge of the shape of the longer 
wavelength image will permit a more accurate characterization of 
light-bending effects in the actual image.  Ultimately, these images
will provide one of the strongest tests of the existence and 
characterization of black holes.

\acknowledgements The National Radio Astronomy Observatory is a facility of the 
National Science Foundation operated under cooperative agreement by Associated 
Universities, Inc. 


\begin{thebibliography}{24}
\expandafter\ifx\csname natexlab\endcsname\relax\def\natexlab#1{#1}\fi

\bibitem[{{Backer}(1978)}]{1978ApJ...222L...9B}
{Backer}, D.~C. 1978, \apjl, 222, L9

\bibitem[{{Balick} \& {Brown}(1974)}]{1974ApJ...194..265B}
{Balick}, B. \& {Brown}, R.~L. 1974, \apj, 194, 265

\bibitem[{{Bower} {et~al.}(2001){Bower}, {Backer}, \&
  {Sramek}}]{2001ApJ...558..127B}
{Bower}, G.~C., {Backer}, D.~C., \& {Sramek}, R.~A. 2001, \apj, 558, 127

\bibitem[{{Bower} {et~al.}(2004){Bower}, {Falcke}, {Herrnstein}, {Zhao},
  {Goss}, \& {Backer}}]{2004Sci...304..704}
{Bower}, G.~C., {Falcke}, H., {Herrnstein}, R.~M., {Zhao}, J., {Goss}, W.~M.,
  \& {Backer}, D.~C. 2004, \sci, 304, 704

\bibitem[{{Bower} {et~al.}(2005){Bower}, {Roberts}, {Yusef-Zadeh}, {Backer},
  {Cotton}, {Goss}, {Lang}, \& {Lithwick}}]{2005ApJ...633..218B}
{Bower}, G.~C., {Roberts}, D.~A., {Yusef-Zadeh}, F., {Backer}, D.~C., {Cotton},
  W.~D., {Goss}, W.~M., {Lang}, C.~C., \& {Lithwick}, Y. 2005, \apj, 633, 218

\bibitem[{{Broderick} \& {Loeb}(2005)}]{2005MNRAS.363..353B}
{Broderick}, A.~E. \& {Loeb}, A. 2005, \mnras, 363, 353

\bibitem[{{Eisenhauer} {et~al.}(2003){Eisenhauer}, {Sch{\"o}del}, {Genzel},
  {Ott}, {Tecza}, {Abuter}, {Eckart}, \& {Alexander}}]{2003ApJ...597L.121E}
{Eisenhauer}, F., {Sch{\"o}del}, R., {Genzel}, R., {Ott}, T., {Tecza}, M.,
  {Abuter}, R., {Eckart}, A., \& {Alexander}, T. 2003, \apjl, 597, L121

\bibitem[{{Falcke} {et~al.}(1993){Falcke}, {Mannheim}, \&
  {Biermann}}]{1993A&A...278L...1F}
{Falcke}, H., {Mannheim}, K., \& {Biermann}, P.~L. 1993, \aap, 278, L1

\bibitem[{{Falcke} \& {Markoff}(2000)}]{2000A&A...362..113F}
{Falcke}, H. \& {Markoff}, S. 2000, \aap, 362, 113

\bibitem[{{Falcke} {et~al.}(2000){Falcke}, {Melia}, \&
  {Agol}}]{2000ApJ...528L..13F}
{Falcke}, H., {Melia}, F., \& {Agol}, E. 2000, \apjl, 528, L13

\bibitem[{{Ghez} {et~al.}(2005){Ghez}, {Salim}, {Hornstein}, {Tanner}, {Lu},
  {Morris}, {Becklin}, \& {Duch{\^e}ne}}]{2005ApJ...620..744G}
{Ghez}, A.~M., {Salim}, S., {Hornstein}, S.~D., {Tanner}, A., {Lu}, J.~R.,
  {Morris}, M., {Becklin}, E.~E., \& {Duch{\^e}ne}, G. 2005, \apj, 620, 744

\bibitem[{{Goldreich} \& {Sridhar}(2006)}]{2006ApJ...640L.159G}
{Goldreich}, P. \& {Sridhar}, S. 2006, \apjl, 640, L159

\bibitem[{{Konigl}(1981)}]{1981ApJ...243..700K}
{Konigl}, A. 1981, \apj, 243, 700

\bibitem[{{Lazio} \& {Cordes}(1998)}]{1998ApJ...505..715L}
{Lazio}, T. J.~W. \& {Cordes}, J.~M. 1998, \apj, 505, 715

\bibitem[{{Melia}(1994)}]{1994ApJ...426..577M}
{Melia}, F. 1994, \apj, 426, 577

\bibitem[{{Narayan} \& {Goodman}(1989)}]{1989MNRAS.238..963N}
{Narayan}, R. \& {Goodman}, J. 1989, \mnras, 238, 963

\bibitem[{{Narayan} {et~al.}(1998){Narayan}, {Mahadevan}, {Grindlay}, {Popham},
  \& {Gammie}}]{1998ApJ...492..554N}
{Narayan}, R., {Mahadevan}, R., {Grindlay}, J.~E., {Popham}, R.~G., \&
  {Gammie}, C. 1998, \apj, 492, 554

\bibitem[{{Quataert} \& {Gruzinov}(2000)}]{2000ApJ...539..809Q}
{Quataert}, E. \& {Gruzinov}, A. 2000, \apj, 539, 809

\bibitem[{{Reid} \& {Brunthaler}(2004)}]{2004ApJ...616..872R}
{Reid}, M.~J. \& {Brunthaler}, A. 2004, \apj, 616, 872

\bibitem[{{Shen} {et~al.}(2005){Shen}, {Lo}, {Liang}, {Ho}, \&
  {Zhao}}]{2005Natur.438...62S}
{Shen}, Z.-Q., {Lo}, K.~Y., {Liang}, M.-C., {Ho}, P.~T.~P., \& {Zhao}, J.-H.
  2005, \nat, 438, 62

\bibitem[{{Trotter} {et~al.}(1998){Trotter}, {Moran}, \&
  {Rodriguez}}]{1998ApJ...493..666T}
{Trotter}, A.~S., {Moran}, J.~M., \& {Rodriguez}, L.~F. 1998, \apj, 493, 666

\bibitem[{{Wilkinson} {et~al.}(1994){Wilkinson}, {Narayan}, \&
  {Spencer}}]{1994MNRAS.269...67W}
{Wilkinson}, P.~N., {Narayan}, R., \& {Spencer}, R.~E. 1994, \mnras, 269, 67

\bibitem[{{Yuan} {et~al.}(2002){Yuan}, {Markoff}, \&
  {Falcke}}]{2002A&A...383..854Y}
{Yuan}, F., {Markoff}, S., \& {Falcke}, H. 2002, \aap, 383, 854

\bibitem[{{Yuan} {et~al.}(2006){Yuan}, {Shen}, \&
  {Huang}}]{2006ApJ...642L..45Y}
{Yuan}, F., {Shen}, Z.-Q., \& {Huang}, L. 2006, \apjl, 642, L45

\end{thebibliography}

\plotone{f1a.eps}
\plotone{f1b.eps}
\plotone{f1c.eps}
\figcaption{({\em left}) Image of Sgr A* at 17.5 cm from the 
VLA+PT observations.  Sgr A* is the
bright source at the center of the image.  The source to the South of Sgr A* is the transient.
The synthesized beam size is indicated in the lower part of the image.  
Contours begin at 5 mJy/beam
and increase by factor of $\sqrt(2)$ to half the peak intensity of Sgr A*.
The synthesized beam size is $1.69 \times 0.56$ arcsec. ({\em center})
Residual image of Sgr A* after subtraction of best-fit point source.  This
shows that the source is clearly resolved East-West and at best marginally
resolved North-South.  ({\em right}) Residual image of Sgr A* after 
subtraction of best-fit Gaussian source.  The source is well-modeled 
to an rms of 2.5 mJy/beam.  \label{fig:vlapt}}

\plotone{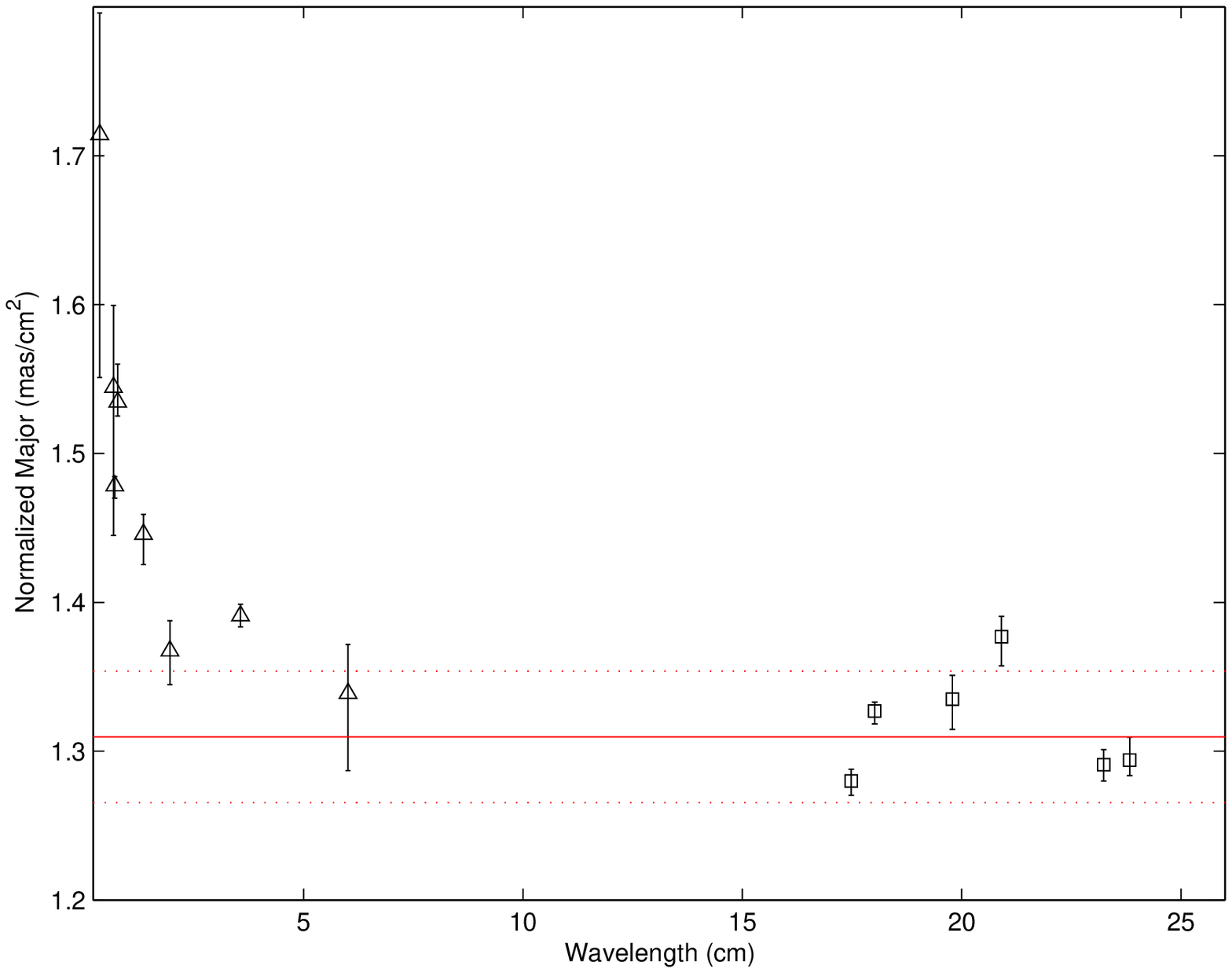}
\figcaption{Measured major axis size
as a function of wavelength.  Triangles are VLBA measurements determined through closure 
amplitude analysis from 
\citet{2004Sci...304..704} 
and 
\citet{2005Natur.438...62S}.
Squares are the new VLA+PT measurements.  The major axis sizes have
been normalized by $\lambda^2$.  The solid red line is
the best-fit scattering value determined from the VLA+PT data alone.  Dotted red lines are $\pm 3\sigma$
of the best-fit scattering law.
\label{fig:results}}

\plotone{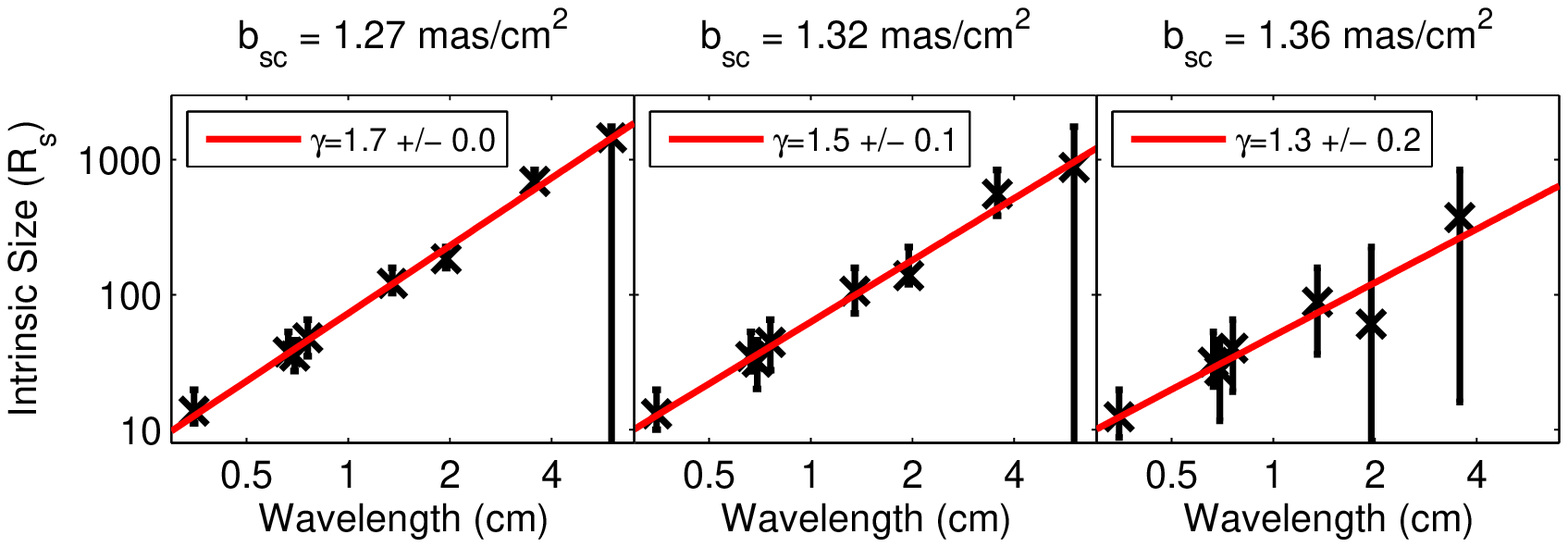}
\figcaption{Intrinsic size in Schwarzschild radii for the East-West dimension 
using three different estimates of the 
major axis scattering law.  We assume a $4\times 10^6 M_\sun$ black hole at a distance of 8 kpc.  In the central panel, we show results for the best-fit 
scattering law.
In the left and right panels, we show the results for the $-3\sigma$ and $+3\sigma$, respectively.
The solid red
lines are the best fit curves for size $\propto \lambda^\gamma$.
\label{fig:intrinsic}}

\begin{deluxetable}{rlrrrrrrr}
\tabletypesize{\scriptsize}
\tablecaption{Apparent and Deconvolved Sizes of Sagittarius A* from VLA+PT Observations of October 2004
\label{tab:decon}}
\tablehead{ 
                      &                       & \multicolumn{3}{c}{Apparent Size} & \multicolumn{3}{c}{Deconvolved Size} & \\
\colhead{ $\lambda$ } & \colhead{Synth. Beam} & \colhead{ $b_{maj}$ } & \colhead{ $b_{min}$ } &\colhead{PA} & \colhead{ $b_{maj}$ } & \colhead{ $b_{min}$ } & \colhead{PA} & \colhead{$S$} \\
\colhead{ (cm) } & \colhead{(mas $\times$ mas,deg)} & \colhead{ (mas) } & \colhead{ (mas) } & \colhead{ (deg) } & \colhead{ (mas) } & \colhead{ (mas) } & \colhead{ (deg) } &\colhead{ (mJy)}}
\startdata
23.8 & $2357 \times 979$, 0.1 & $2433.5^{+  8.2}_{-  8.2}$ & $1222.6^{+  5.2}_{-  3.1}$ & $ -0.21^{+  0.14}_{-  0.14}$ & $734.7^{+  8.7}_{-  5.9}$ & $602.4^{+ 36.0}_{- 29.9}$ & $97.7^{+ 1.1}_{- 7.5}$ & $471.9^{+  1.6}_{-  1.6}$ \\ 
23.2 & $2191 \times 714$, 4.8 & $2252.1^{+  8.4}_{-  5.1}$ & $985.5^{+  3.0}_{-  3.7}$ & $  5.83^{+  0.10}_{-  0.10}$ & $697.1^{+  5.4}_{-  5.9}$ & $496.3^{+ 40.1}_{- 19.6}$ & $76.5^{+ 2.7}_{- 2.1}$ & $490.6^{+  1.5}_{-  1.9}$ \\ 
20.9 & $1996 \times 646$, 7.6 & $2049.4^{+  6.1}_{-  6.1}$ & $849.2^{+  2.5}_{-  2.5}$ & $  9.06^{+  0.07}_{-  0.07}$ & $601.8^{+  6.0}_{-  8.6}$ & $397.3^{+ 27.2}_{- 19.1}$ & $66.1^{+ 3.9}_{- 2.9}$ & $432.1^{+  1.3}_{-  1.3}$ \\ 
19.8 & $1840 \times 581$, 10.4 & $1877.8^{+  5.4}_{-  5.4}$ & $751.1^{+  2.2}_{-  2.2}$ & $ 11.92^{+  0.06}_{-  0.06}$ & $522.8^{+  6.3}_{-  8.0}$ & $306.6^{+ 31.2}_{- 21.1}$ & $70.5^{+ 3.7}_{- 2.7}$ & $441.5^{+  1.3}_{-  1.3}$ \\ 
18.0 & $1737 \times 564$, 7.2 & $1758.6^{+  2.5}_{-  2.5}$ & $695.9^{+  1.0}_{-  1.0}$ & $  8.23^{+  0.06}_{-  0.06}$ & $430.8^{+  1.9}_{-  2.8}$ & $236.7^{+ 18.6}_{- 14.4}$ & $74.9^{+ 1.8}_{- 1.3}$ & $505.0^{+  0.7}_{-  0.7}$ \\ 
17.5 & $1689 \times 556$, 6.8 & $1706.4^{+  2.5}_{-  2.5}$ & $667.6^{+  1.0}_{-  1.0}$ & $  7.71^{+  0.06}_{-  0.06}$ & $391.2^{+  2.4}_{-  3.0}$ & $206.3^{+ 21.6}_{- 15.5}$ & $74.5^{+ 2.1}_{- 1.5}$ & $511.9^{+  0.7}_{-  0.7}$ \\ 
\enddata
\tablecomments{The quoted errors in the peak flux density $S$ do not include the $\sim 10\%$
error in absolute flux density.}
\end{deluxetable}

\begin{deluxetable}{rrrr}
\tablecaption{Intrinsic Size of Sagittarius A*
\label{tab:intrinsic}}
\tablehead{ 
		      & \colhead{$b_{sc}=1.26 {\rm\ mas/cm^2}$} &\colhead{$b_{sc}=1.31 {\rm\ mas/cm^2}$} &\colhead{$b_{sc}=1.36 {\rm\ mas/cm^2}$} \\
\colhead{ $\lambda$ } & \colhead{ Size }  & \colhead{ Size }  & \colhead{ Size }  \\
\colhead{ (cm) } & \colhead{ (mas) } & \colhead{ (mas) } & \colhead{ (mas) } }
\startdata
 0.35  & $  0.142^{+ 0.062}_{- 0.025}$ & $  0.136^{+ 0.069}_{- 0.032}$& $  0.128^{+ 0.076}_{- 0.040}$\\ 
 0.67  & $  0.395^{+ 0.157}_{- 0.047}$ & $  0.362^{+ 0.190}_{- 0.080}$& $  0.324^{+ 0.228}_{- 0.118}$\\ 
 0.69  & $  0.373^{+ 0.110}_{- 0.077}$ & $  0.331^{+ 0.152}_{- 0.120}$& $  0.280^{+ 0.203}_{- 0.170}$\\ 
 0.76  & $  0.505^{+ 0.177}_{- 0.128}$ & $  0.461^{+ 0.221}_{- 0.172}$& $  0.410^{+ 0.272}_{- 0.223}$\\ 
 1.35  & $  1.295^{+ 0.369}_{- 0.174}$ & $  1.117^{+ 0.547}_{- 0.351}$& $  0.897^{+ 0.767}_{- 0.571}$\\ 
 1.95  & $  2.028^{+ 0.408}_{- 0.304}$ & $  1.500^{+ 0.936}_{- 0.223}$& $  0.563^{+ 1.873}_{- 0.562}$\\ 
 3.56  & $  7.491^{+ 1.524}_{- 0.281}$ & $  5.953^{+ 3.062}_{- 1.819}$& $  3.735^{+ 5.280}_{- 3.732}$\\ 
 6.01  & $ 16.427^{+ 2.981}_{-16.410}$ & $ 10.094^{+ 9.314}_{-10.083}$& $  0.000^{+19.408}_{- 0.000}$\\ 
\enddata
\tablecomments{The intrinsic size is computed for the best-fit value to the 
major axis scattering size ($b_{sc}=1.31 {\rm\ mas/cm^2}$) and for 
$\pm 3\sigma$ values.}
\end{deluxetable}

\end{document}